\documentclass[sigconf]{acmart}
\usepackage{tabularx}
\usepackage{mdwlist}
\usepackage{subcaption}
\hypersetup{colorlinks=false,linkcolor=blue,urlcolor=blue,citecolor=red}
\usepackage[labelfont=bf,textfont=bf]{caption}
\usepackage{multirow}
\usepackage{array}
\usepackage[linesnumbered,ruled]{algorithm2e}
\SetKwComment{Comment}{$\triangleright$\ }{}
\usepackage{xcolor}
\usepackage{epstopdf}
\usepackage{enumitem}
\usepackage[utf8]{inputenc} 
\usepackage[T1]{fontenc}    
\usepackage{hyperref}       
\usepackage{url}            
\usepackage{booktabs}       
\usepackage{amsfonts}       
\usepackage{nicefrac}       
\usepackage{microtype}      
\usepackage{xcolor}         
\usepackage{enumitem}
\usepackage{listings}
\usepackage{graphicx}  
\usepackage{float}  
\usepackage{subcaption} 
\usepackage{placeins}
\usepackage{amsmath}
\usepackage{adjustbox}
\usepackage[utf8]{inputenc}

\copyrightyear{2025}
\acmYear{2025}
\setcopyright{acmlicensed}\acmConference[WWW Companion '25]{Companion Proceedings of the ACM Web Conference 2025}{April 28-May 2, 2025}{Sydney, NSW, Australia}
\acmBooktitle{Companion Proceedings of the ACM Web Conference 2025 (WWW Companion '25), April 28-May 2, 2025, Sydney, NSW, Australia} \acmDOI{10.1145/3701716.3715232} \acmISBN{979-8-4007-1331-6/2025/04}

\begin{document}


\title{HedgeAgents: A Balanced-aware Multi-agent Financial Trading System}

\author{Xiangyu Li}
\authornote{Both authors contributed equally to this paper}
\email{65603605lxy@gmail.com}
\orcid{0009-0002-8261475r}
\affiliation{
  \institution{South China University of Technology}
  \city{Guangzhou}
  \country{China}
}

\author{Yawen Zeng}
\orcid{0000-0003-1908-1157}
\email{yawenzeng11@gmail.com}
\authornotemark[1]
\affiliation{%
  \institution{ByteDance}
  \city{Beijing}
  \country{China}
}

\author{Xiaofen Xing}
\email{xfxing@scut.edu.cn}
\orcid{0000-0002-0016-9055}
\affiliation{
  \institution{South China University of Technology}
  \city{Guangzhou}
  \country{China}
}

\author{Jin Xu}
\orcid{0009-0001-8735-3532}
\email{jinxu@scut.edu.cn}
\affiliation{%
  \institution{South China University of Technology}
  \institution{Pazhou Lab}
  \city{Guangzhou}
  \country{China}
}

\author{Xiangmin Xu}
\authornote{Corresponding author}
\orcid{0009-0001-8735-3532}
\email{xmxu@scut.edu.cn}
\affiliation{%
  \institution{South China University of Technology}
  \city{Guangzhou}
  \country{China}
}

\hyphenpenalty=2000
\tolerance=6000

\ccsdesc[500]{Information systems~Financial Trading Systems}

\begin{abstract}
As automated trading gains traction in the financial market, algorithmic investment strategies are increasingly prominent. While Large Language Models (LLMs) and Agent-based models exhibit promising potential in real-time market analysis and trading decisions, they still experience a significant -20\% loss when confronted with rapid declines or frequent fluctuations, impeding their practical application. Hence, there is an imperative to explore a more robust and resilient framework. This paper introduces an innovative multi-agent system, \textbf{HedgeAgents}, aimed at bolstering system robustness via ``hedging'' strategies. In this well-balanced system, an array of hedging agents has been tailored, where HedgeAgents consist of a central fund manager and multiple hedging experts specializing in various financial asset classes. These agents leverage LLMs' cognitive capabilities to make decisions and coordinate through three types of conferences. Benefiting from the powerful understanding of LLMs, our HedgeAgents attained a 70\% annualized return and a 400\% total return  over a period of 3 years. Moreover, we have observed with delight that HedgeAgents can even formulate investment experience comparable to those of human experts (\url{https://hedgeagents.github.io/}).
\end{abstract}

\keywords{Quantization Finance, Large Language Model, Multi-agent Systems}

\maketitle

\begin{figure}[h]
    \center
    \vspace{-0.8cm}
    \includegraphics[width=0.38\textwidth]{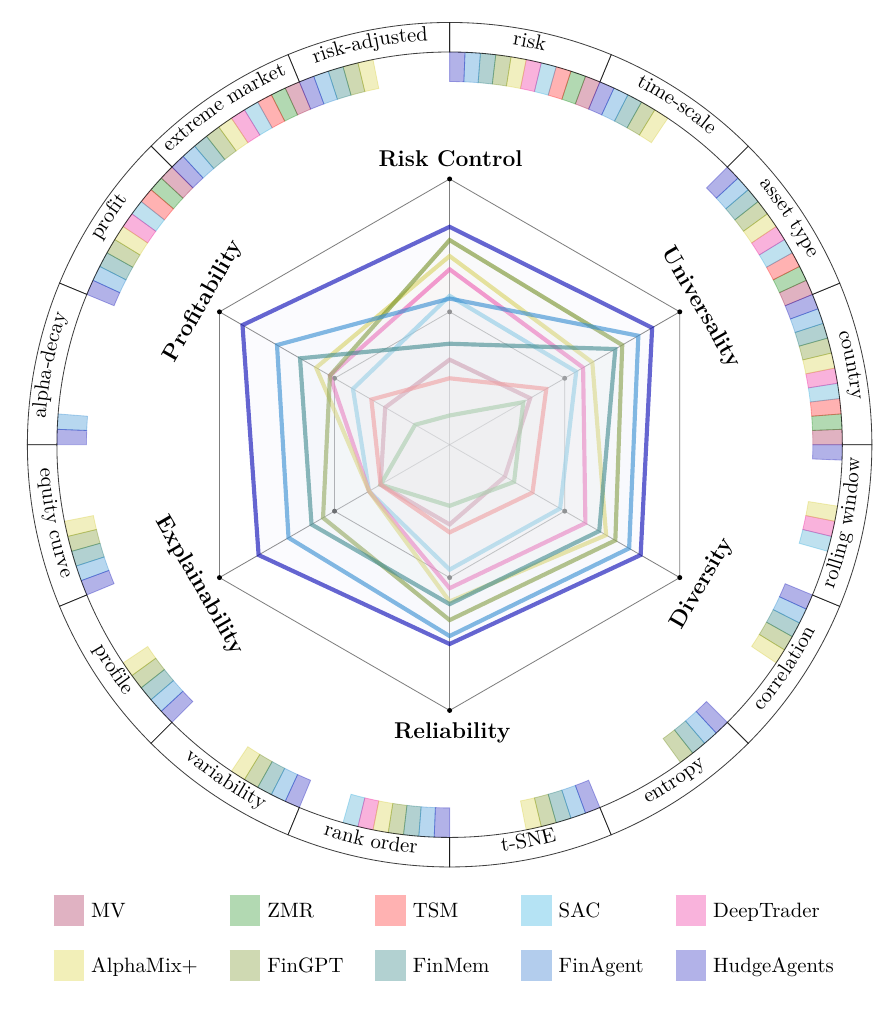}
    \vspace{-0.5cm}
    \caption{Our method outperformed all baselines on the PRUDEX benchmark \cite{sun2023prudexcompass} across six dimensions. Marks on the inner circle represent the market average, while the outer layer details the measures evaluated.
    }
    \label{fig:compass}
    \vspace{-0.5cm}
\end{figure}


\section{Introduction}

\begin{figure*}[htbp]
  \centering
   \includegraphics[width=0.99\textwidth,keepaspectratio]{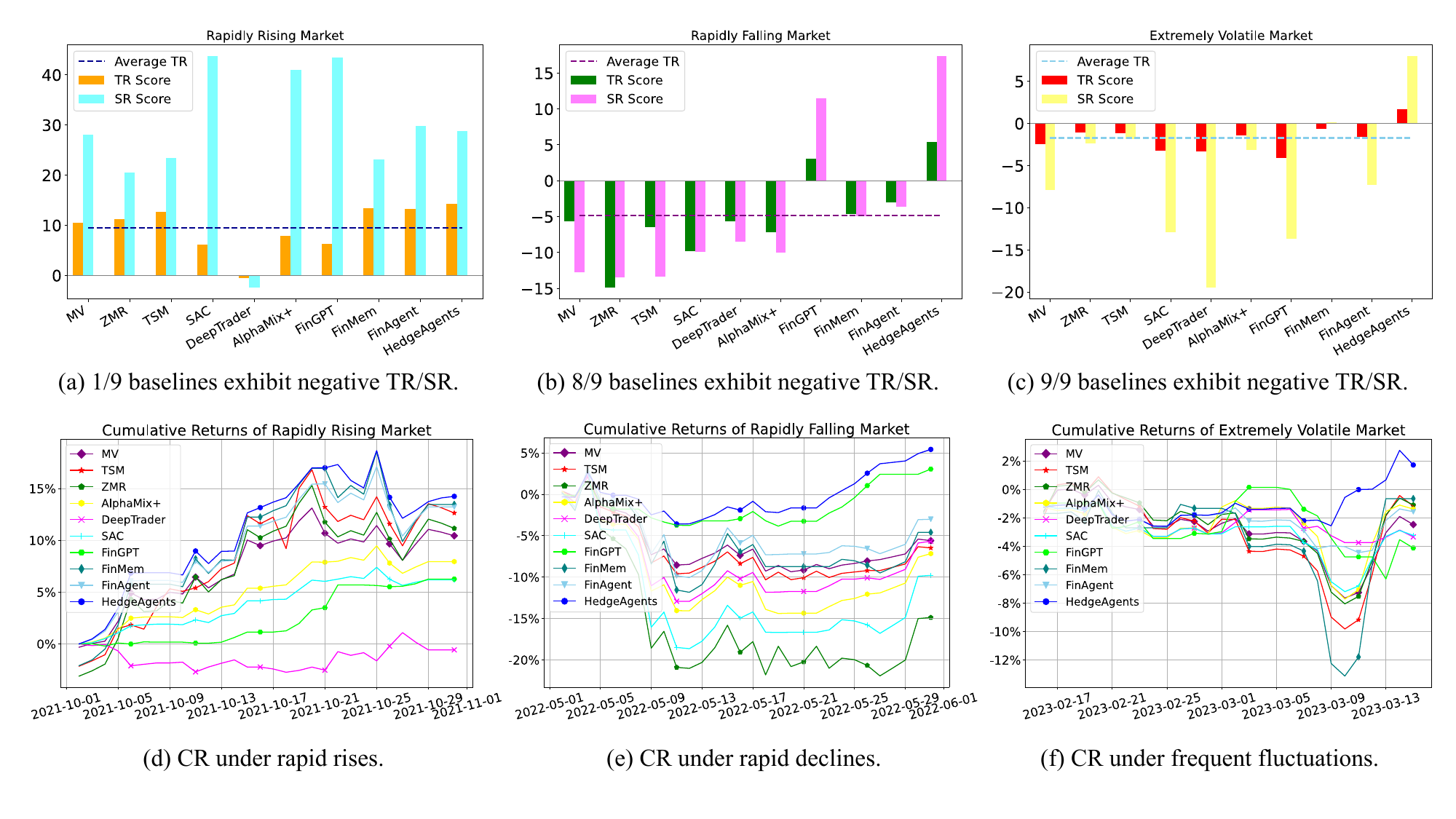}
  \vspace{-0.6cm}
  \caption{Performance of state-of-the-art models on Total Return (TR), Sharpe Ratio (SR) and Cumulative Returns (CR) in real-world scenarios, with the majority exhibiting negative scores.}
  \label{fig:intro1}
  \vspace{-10pt}
\end{figure*}
In contemporary financial markets, automated trading has emerged as a pivotal investment strategy \cite{bankivestment}. Leveraging advanced algorithms, traders can actively track market trends in real-time, enabling swift and precise decision-making, thereby enhancing investment yields while mitigating risk \cite{kanamura2021pricing}. The integration of artificial intelligence into financial transactions has unlocked significant opportunities, notably with the advent of large language models (LLMs) \cite{touvron2023llama,chowdhery2023palm,zeng2021multi}. These cutting-edge models possess the capability to analyze extensive financial and news information, forecast market variations, and even produce financial analyses and reports \cite{finreportacm}. Moreover, LLM-based agents can refine trading strategies through the simulation of market dynamics. 

Despite the impressive performance of these cutting-edge models \cite{yang2023autogpt,pan2023llms}, they still lack the robustness required to withstand real-world fluctuations. As depicted in Figure \ref{fig:intro1}, the performance of state-of-the-art models is notably poor on Total Return (TR) and Sharpe Ratio (SR) metrics. For instance, in ``rapid decline'' scenarios, 8 out of 9 baselines exhibit negative scores below zero. Meanwhile, both RL-based DeepTrader \cite{wang2021deeptrader} and LLM-based FinGPT \cite{yang2023fingpt} experience financial losses ranging from {\color{red}-15\% to -20\%} in ``frequent fluctuations'' scenarios.
This limitation stems from the absence of risk management mechanisms within these models or systems, impeding their practical application in real markets. Compounded by the inherent complexity and uncertainty already prevalent in financial markets, the challenge of designing and optimizing trading strategies is further exacerbated.

In this paper, our aim is to incorporate the concept of ``hedging'' in the development of a robust and reliable financial trading system. Empirically, hedging serves as a strategic tool that enables traders to sustain returns amidst market volatility by constructing diversified portfolios. Along this line, we are committed to building a multi-agent system that mirrors real-world market, with a LLM acting as the brain \cite{hong2023metagpt,wu2023bloomberggpt}. A key challenge lies in the configuration and coordination of these hedge agents, which constitutes our primary focus. Illustrated in Figure \ref{fig:intro1}, unconstrained agent system (e.g. FinAgent \cite{zhang2024multimodal}) can only achieve suboptimal performance.

We are devoted to developing a multi-agent hedging system named \textbf{HedgeAgents}. Within this well-balanced framework, we have established a comprehensive array of hedge agents, comprising a fund manager and a team of specialized experts in various fields such as Stocks, Forex, and Bitcoin. Each expert is tasked with overseeing their designated area, while the fund manager plays a pivotal role in orchestrating their efforts through facilitating discussions, conducting reviews, and consolidating insights. Leveraging the profound understanding capabilities of LLM, these agents (from their memories) have distilled invaluable investment experience akin to those crafted by human counterparts! Drawing from these rich experiences and hedging expertise, our HedgeAgents not only demonstrates unwavering stability even in the face of extreme conditions (Figure \ref{fig:intro1}), but also wins across all metrics, achieving a total return of {\color{teal}400\% over 3 years} (Figure \ref{fig:compass}). We hold a strong belief that this research will pave the way for a dependable automated system that enhances the security of financial investments. 

The contributions are summarized as follows: 
\begin{itemize}[leftmargin=*]
\item {}To the best of our knowledge, this work represents a pioneering effort in integrating ``hedging'' within a multi-agent environment to develop a robust and reliable financial trading system.
\item{} We have established a hedging portfolio consisting of three experts and one manager, collaborating through three types of conference, with a LLM serving as the central cognitive hub.
\item{} Extensive experiments has demonstrated that our framework delivers optimal performance on all metrics, achieving a 70\% annualized return and a total return of 400\% over 3 years. Furthermore, the hedge experts within our framework have produced an invaluable investment experience in their memories akin to human expertise. The codes are available at here\footnote{https://hedgeagents.github.io/}.
\end{itemize}

\begin{figure*}[htbp]
  \centering
   \includegraphics[width=0.95\textwidth,keepaspectratio]{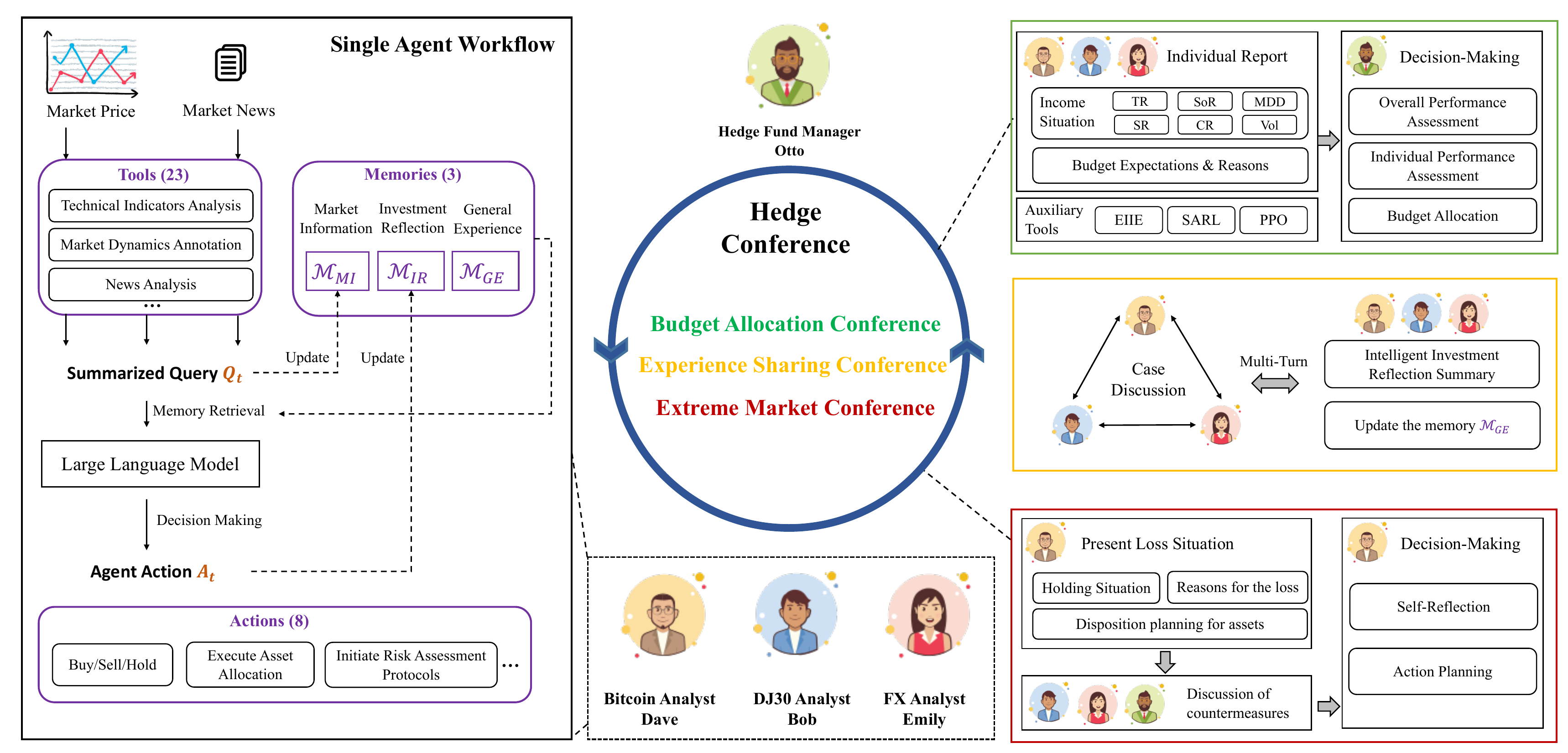}
  \vspace{-0.4cm}
  \caption{Our HedgeAgents comprise 3 hedging agents and 1 manager. Each agent is equipped with 23 tools and possesses 3 types of memory to execute 8 actions. Furthermore, collaboration among multiple agents can be categorized into three types of conferences: {\color{teal}budget allocation}, {\color{olive}experience sharing}, and {\color{red}extreme market conference}.}
  \label{fig:framework}
  \vspace{-0.4cm}
\end{figure*}

\section{Related Work}
\subsection{Quantitative Finance}
Quantitative finance is an interdisciplinary field that merges finance with mathematical and statistical methods to address complex financial challenges \cite{kou2019machine,kanamura2021pricing}. It plays a crucial role in the valuation of sophisticated financial derivatives, portfolio optimization, and the analysis of market dynamics \cite{tavella2003quantitative,horvath2021deep}. With the advent of machine learning, the field has seen a surge in predictive modeling capabilities, enhancing both the accuracy of market forecasts and the efficiency of algorithmic trading systems \cite{mieg2022volatility}, such as MV \cite{YU2011367}, DeepTrader \cite{wang2021deeptrader}. As financial markets continue to evolve, the integration of quantitative techniques is essential for developing effective strategies that can withstand market uncertainties \cite{finreportacm}.

\subsection{LLM-based Agent}
LLM-based Agents, leveraging the cognitive and generative prowess of Large Language Models \cite{touvron2023llama, chowdhery2023palm}, promising advancements towards Artificial General Intelligence (AGI)\cite{yang2023autogpt}. LLMs play a vital role in enhancing the autonomy\cite{Wang_2024}, reactivity, and social interaction capabilities of agents. This enables agents to perform a range of complex tasks, including natural language interaction, knowledge integration, information memory, logical reasoning, and strategic planning\cite{sumers2023cognitive, park2023generative} . LLM-based agent systems based on LLMs hold promise for impactful contributions across sectors like finance (e.g. FinGPT \cite{yang2023fingpt}, FinAgent \cite{zhang2024multimodal}), offering novel approaches and sophisticated solutions to intricate challenges \cite{hong2023metagpt,wu2023bloomberggpt,zhang2023appagent}.

\section{Proposed Method}

\subsection{Preliminaries}
The objective is to optimize returns while minimizing risk using hedging strategies. Specifically, this task involves utilizing financial data, encompassing prices and news, as input. Our framework then generates and implements trading actions, such as buying and selling, across three domains: Bitcoin, Stocks, and Forex.

\subsection{Overall Framework}
The HedgeAgents framework simulates the architecture of a real hedge fund company, aiming to optimize risk hedging for multi-asset investment portfolios. As depicted in Figure \ref{fig:framework}, the framework comprises four agents: Bitcoin Analyst Dave, Dow Jones Analyst Bob, Forex Analyst Emily, and Hedge Fund Manager Otto. Each of the three analysts is in charge of managing a specific asset. Otto, as their supervisor, is responsible for the overall risk management of the investment portfolio and the distribution of the asset investment budget.
To achieve effective collaboration, we have established three types of multi-agent coordination meetings: Budget Allocation Conference (BAC), Experience Sharing Conference (ESC) and Extreme Market Conference (EMC). These conferences serve to facilitate budget allocation, experience summary, and swift emergency actions.

\subsection{Definitions of Single Agent}
In this section, we will provide a comprehensive overview of the composition and execution process of a single agent, designed to simulate the human decision-making process in investments. Each agent comprises a range of financial analysis tools $\mathcal{T}$, along with definitions for profile, action $\mathcal{A}$, memory and reflection $\mathcal{M}$. Among them, there are 23 tools, encompassing Indicator Analysis, Cryptocurrency Market Analysis, and Risk Management. Actions comprise 8 types, such as Buy/Sell/Hold for current assets. Memory is categorized into 3 types: basic Market Information Memory $\mathcal{M}_{MI}$, Investment Reflection $\mathcal{M}_{IR}$, and General Experience $\mathcal{M}_{GE}$.

\begin{center}
\fcolorbox{black}{gray!10}{\parbox{1\linewidth}{
<Simplified Prompt Template> \\
You are \{Dave Profile\}. The market environment today includes \{Prices\}, \{News\}. Through financial analysis tools, \{Tool Results\} can be obtained. The output format should be JSON, such as \{Examples\}.
}}
\end{center}


The following will introduce the workflow for a single agent. We have integrated the LLM-based intelligent investment agent into a reinforcement learning framework through a reflection-driven decision-making process, enabling the flexible definition of the reasoning processes. The reflection-driven process includes three steps: memory retrieval, decision making, and reflection update.

\textbf{Memory Retrieval.} We employ retrieval strategy to search for reliable experiences to enhance decision-making. Notably, due to the substantial input received by each agent at time step $t$ (comprising market information, financial news, and tool results), we will first compile a summarized query $Q_t$ and then utilize it to retrieve K=5 similar cases $\mathcal{M}_{ret}$ from all memories $\mathcal{M}=\{\mathcal{M}_{MI}, \mathcal{M}_{GE}, \mathcal{M}_{IR}\}$.

\textbf{Decision Making.} Based on the experience $\mathcal{M}_{ret}$, we extend reinforcement learning as follows:
\begin{equation}
    \begin{aligned}
	\pi_{{\theta}^*}^{}=\arg\max_{{\pi}_{\theta}}{\mathbb{E}           {_{\pi}}_{\theta}}\left[\sum_{t=0}^T\gamma r_{t}\left|s_t=s,\mu_t=\mu \right.\right],
    \end{aligned}
\end{equation}
where $r_t$ is the reward at time step $t$, which depends on the environment state $s_t$ (e.g. market indicators) and action $a_t$ (e.g. bug/sell). An action $a_t$ is determined through the following process:
\begin{equation}\label{eqn:2}
    \begin{aligned}
        {\pi}_{Agent}\left(a_t|s_t,\mu_t\right) & \equiv {\mathcal{D}}\left(LLM\left(\phi_D\left(s_t,\mathcal{M}_{ret}\right)\right)\right) , \\
    \end{aligned}
\end{equation} 
where the prompt template $\phi_D(\cdot)$ is designed to receive responses via a LLM, which are then processed by parsing function $D(\cdot)$ to determine compatible actions in this environment. Therefore, the goal of a agent is to find the policy $\mu$ to optimize the total return.

\textbf{Reflection Update.} The basic market information and summarized queries $Q_t$ will be stored in the basic memory $\mathcal{M}_{MI}$, while the reflection and actions during decision-making will be updated in the investment reflection $\mathcal{M}_{IR}$.

\subsection{Coordination of Hedging Agents}
In this section, we design three collaborative meetings to implement the hedging strategies of multiple agents. Among them, the periodic Budget Allocation Conference realizes the trading of multiple assets, with Dave (Bitcoin), Bob (Stocks), Emily (Forex), and manager Otto participating. The Extreme Market Conference serves as an emergency mechanism, enhancing the overall robustness of our system, while the Experience Sharing Conference fosters the exchange of valuable investment insights.

\subsubsection{Budget Allocation Conference}
This conference spans a 30-day cycle, wherein Dave, Bob, and Emily present reports to Manager Otto aimed at budget allocation. Each report encompasses the current profit situation $R_{p}$ alongside budget expectations \& reasons $R_{b}$. Subsequently, manager Otto consolidates all reports and auxiliary tools $R_{a}$ via a prompt template ${\phi}_{\rho}$, to evaluate expected future returns ${\rho}_{i}$.
\begin{equation}
    \begin{aligned}
        {\rho}_{i}=LLM\left({\phi}_{\rho}\left(R_{p},R_{b},R_{a}\right)\right), i \in \{Dave, Bob, Emily\} .
    \end{aligned}
\end{equation}
Furthermore, to refine the asset portfolio weighting, three additional indicators are introduced: expected total return $I_{etr}$, overall portfolio risk $I_{pr}$, and conditional expected draw down risk $I_{cvar}$.
\begin{equation}
    I_{etr} = \sum_{i=1}^{3} \omega_{i} \rho_{i}, \quad
    {I_{pr}^2} = \sum_{i=1}^{3} \sum_{j=1}^{3} \omega_i \omega_j I_{ij}, \quad
    I_{cvar} = \frac{1}{1-\alpha} \int_{1}^{\alpha} \mathrm{VaR}(p) \, dp ,
\end{equation}
where $w_i$ represents the weight of asset $i$, $\sigma_{ij}$ is the covariance between the returns of asset $i$ and asset $j$, and $\mathrm{VaR}(\alpha)$ is the Value at Risk at confidence level $\alpha$, estimated using the historical simulation method.
Finally, budget allocations are determined via an optimization objective function:
\begin{equation}
\begin{aligned}
    \omega^* & =\arg\max_{\omega}I_{etr}-\lambda_{1}{I_{pr}}-\lambda_{2}I_{cvar},\quad
    \text{s.t.} \underset{}{\overset{}{\underset{i}{\overset{}{\sum}}{\omega}_i=1,{\omega}_i\ge0}} , 
\end{aligned}
\end{equation}
where $\lambda_1$ and $\lambda_2$ are risk aversion coefficients, which ensure the sum of the allocated budget weights is 1 and non-negative.

\subsubsection{Experience Sharing Conference}
This is a regular meeting held at the end of each investment cycle, where 3 hedging agents engage in multi-round knowledge accumulation. In each round, each agent will recapitulate a typical case $C_{t}$ from the memory $\mathcal{M}_{IR}$ for discussion among peers, with all insights $D_{t}$ being archived in memory $\mathcal{M}_{GE}$. In this way, the invaluable experience is consolidated to bolster future decision-making.
\begin{equation}
    \begin{aligned}
    \mathcal{M}_{GE}=LLM({\phi}_{c}(D_{t},C_{t})),
    \end{aligned}
\end{equation}
where the LLM aims to update the general experience memory $\mathcal{M}_{GE}$ via a prompt template $\phi_{c}$.

\subsubsection{Extreme Market Conference}
This is an interim session convened to address volatile market conditions, as identified by manager Otto, characterized by a daily amplitude exceeding 5\% or a cumulative three-day amplitude surpassing 10\%. During this session, the crisis agent facing challenges are required to present their current portfolio holdings, articulate the reasons behind the crisis, and outline the proposed plans. Following this, Otto and other agents will offer suggestions $\mathcal{S}_{B}$, $\mathcal{S}_{C}$, $S_{E}$ aimed at fostering reflection and optimizing the actions, with the ultimate goal of navigating through the turbulent period effectively.
\begin{equation}
    \begin{aligned}
        \mathcal{S}=(1-{\lambda_3}){\phi_B}(\mathcal{S}_{B},\mathcal{S}_{C})+{\lambda_3}{\phi_B}(\mathcal{S}_{E}),
    \end{aligned}
\end{equation}
where ${\lambda_3}$ is used to balance the importance of suggestions $\mathcal{S}$, and $\phi_B(\cdot)$ is a prompt template. The optimization process is similar to Eqn.\ref{eqn:2}.
\begin{equation}
\begin{aligned}
    \pi^* &= \arg\max_{\pi(\cdot),\mu(\cdot)} \mathbb{E} \pi \left[ \sum_{t=0}^T \gamma r_{t} \left| s_t=s, \mu_t=\mu \right. \right] \\
    \text{s.t.} \quad &\pi(a_t|s_t,\mu_t) =\mathcal{D}\left(LLM\left(\phi_B\left(s_t,\mathcal{S}\right)\right)\right)  \quad \forall t
\end{aligned}
\end{equation}

\begin{table*}[]
\caption{Performance comparison of all baselines on 9 evaluation metrics. \textbf{Bold} represents optimal performance, while \underline{underline} represents suboptimal.}
\label{tab:main}
\small
\centering
\normalsize
\setlength{\tabcolsep}{6pt} 
\renewcommand{\arraystretch}{0.99} 
\resizebox{1\textwidth}{!}{ 
\begin{tabular}{cccccccccccl}
\hline
\hline
Categories       & Models           & ARR(\% )& TR(\%)  & SR      & CR      & SoR     & MDD(\%) & Vol(\%) & ENT     & ENB     \\
\midrule
                 & Bitcoin          & 12.92 & 43.97 & 0.54 & 1.19   & 12.49 & 76.63 & 3.40  &    --     &    --     \\
Market Trends           & FX               & 4.08& 12.74& 0.61& 0.93& 11.99& \underline{15.56}&\textbf{0.38}&    --     &    --     \\
                 & DJIA             & 7.64& 24.70& 0.59& 1.06    & 11.72& 21.94& \underline{0.78}&    --     &     --    \\\midrule
                 & MV               & 13.03& 44.39& 0.71& 1.25& 16.14& 32.04& \textcolor{black}{1.13}& 1.09& 1.02\\
Rule-based        & ZMR              & -7.25& -20.21& -0.52& -3.13& -5.15& 61.52& 1.98& 1.55& 1.11\\
                 & TSM              & 19.13& 69.09& 0.78& 1.53& 18.21& 39.14& 1.55& 1.10& 1.09\\\midrule
                 & SAC              & 24.71& 93.94& 1.16& 3.12& 23.15& 21.56& 1.16& 1.62& 1.14\\
RL-based         & DeepTrader       & 32.78& 134.11& 1.41& 4.06& 30.43& 20.95& 1.21& 2.02& 1.30\\
                 & AlphaMix+        & 37.59& 160.47& 1.62& 3.69    & 35.52& 25.56& 1.17& \underline{2.93}& 1.22\\\midrule
                 & FinGPT           & 34.22& 141.82& \underline{1.93}& \underline{7.64}& \underline{39.57}   &\textcolor{black}{ 17.08}& 8.76& 1.76& \textcolor{black}{1.33}\\
LLM-based        & FinMem           & \textcolor{black}{47.67}& \textcolor{black}{221.99}& 1.20& 4.02& 25.42& 32.39& 2.16& 1.99& 1.25\\
                 & FinAgent         & \underline{53.54}& \underline{261.98}  & \textcolor{black}{1.80}&\textcolor{black}{ 4.52}&\textcolor{black}{ 39.12}& 28.24& 1.42& \textcolor{black}{2.85}& \underline{1.41}\\\midrule
Ours             &\textbf{HedgeAgents}     & \textbf{71.60}& \textbf{405.34}& \textbf{2.41}& \textbf{11.02}& \textbf{58.00}& \textbf{14.21}& 1.30& \textbf{3.13}& \textbf{1.53}\\
\midrule
\multicolumn{2}{c}{Improvement(\%)} & 33.75& 54.72& 24.49& 44.28& 46.58& 16.76&  \multicolumn{1}{c}{--}& 6.85& 8.51\\
\hline
\hline
\end{tabular}
}
\end{table*}

\begin{figure}[htbp]
  \centering
   \includegraphics[width=0.49\textwidth,keepaspectratio]{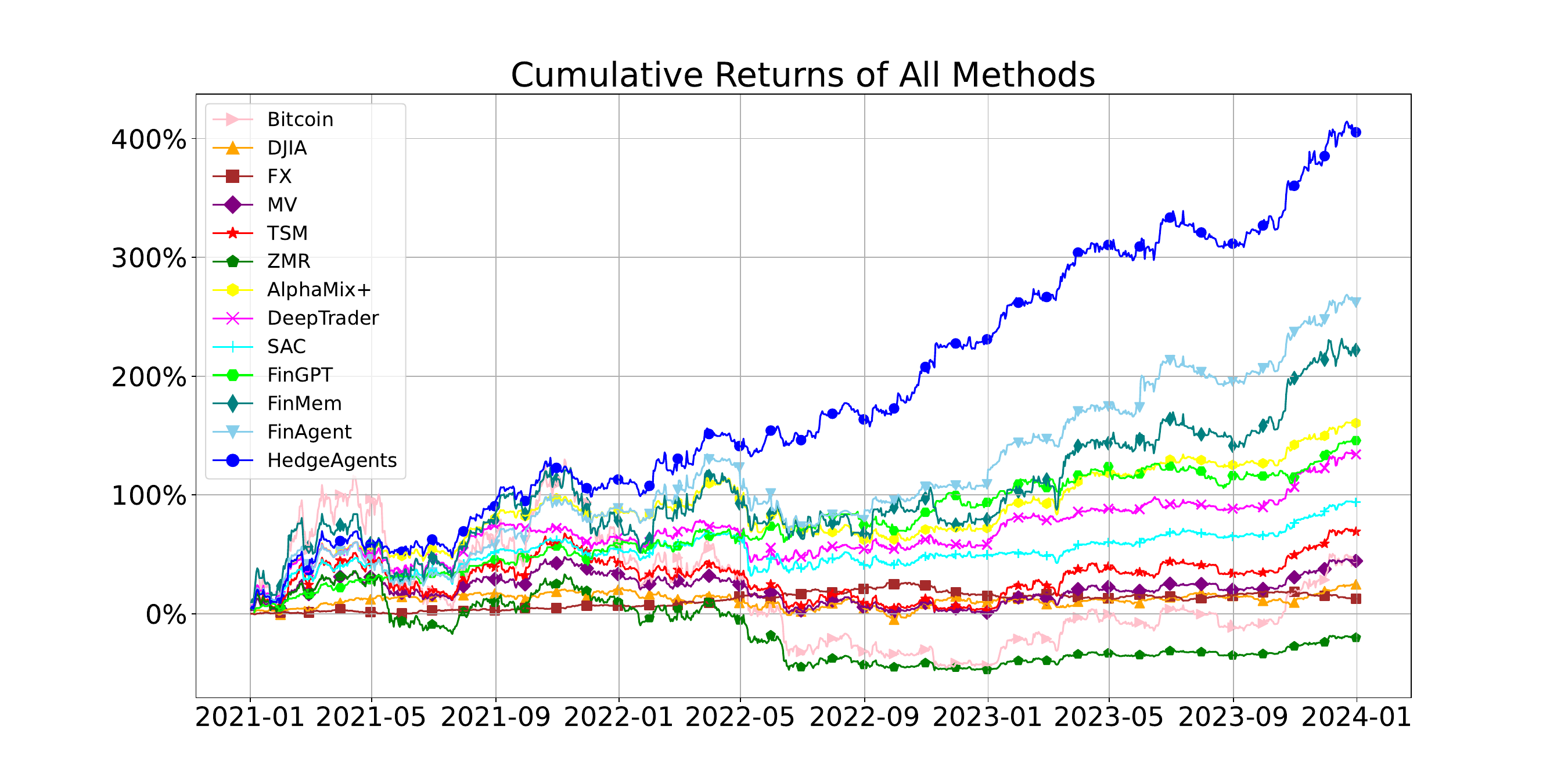}
  \caption{Cumulative Returns Comparison of all baselines and our HedgeAgents.}
  \label{fig:CRAll}
\end{figure}

\section{Experiments}
\subsection{Datasets}\label{sec:datasets}
We have compiled a comprehensive financial dataset comprising Bitcoin, foreign exchange, and the Dow Jones component stocks. These data were sourced from reputable financial databases, namely Yahoo Finance and the Alpaca News API. The dataset spans from January 1, 2015, to December 31, 2023, encompassing daily data points such as open, high, low, and close prices, as well as volume and adjusted close prices. Additionally, daily news updates\footnote{Following \cite{finreportacm}, we only use processed news headlines that cover the key information of the news.} and 60 standard technical analysis indicators are included for each asset.


\subsection{Evaluation Metrics}\label{subsec:eva}
We compare HedgeAgents and baselines in terms of 9 financial metrics following \cite{sun2023trademaster,qin2023earnhft}, which include 2 profit metrics: Total Return (TR), Annual Return Rate (ARR); 3 risk-adjusted profit metrics: Sharpe Ratio (SR), Calmar Ratio (CR), Sortino Ratio (SOR); 2 risk metrics: Maximum Drawdown (MDD), Volatility (VOl); and 2 diversity metrics: Entropy (ENT) and Effect Number of Bets (ENB).


\subsection{Implementation Details}
The dataset is divided using January 1, 2021, as the cutoff point, with data from January 1, 2015, to December 31, 2020, designated as the training set, and data from January 1, 2021, to December 31, 2023, as the testing set. Notably, during the testing phase, only historical prices can be access to avoid data leakage issues. To ensure the fairness of the evaluation results, all baseline models are trained and tested in the same reinforcement learning environment.

For baseline configurations, we employed Optuna \cite{akiba2019optuna} for hyperparameter optimization.
For LLM-based approaches and HedgeAgents, we consistently use the GPT-4-1106-preview version with a temperature setting of 0.7 to balance consistency and creativity. The memory module in our framework is designed as a text similarity-based storage and retrieval mechanism, utilizing the text-embedding-3-large model \cite{openai2023textembedding3}, with a top-k value of 5. We ensured LLM-based methods are adjusted to their optimal configurations as suggested in their respective studies.

\begin{table*}[]
\caption{Ablation analysis on three conference. \checkmark denote the inclusion of components.}
\label{tab:rq2}
\small
\centering
\normalsize
\setlength{\tabcolsep}{6pt} 
\renewcommand{\arraystretch}{0.99} 
\resizebox{1\textwidth}{!}{ 
\begin{tabular}{ccc|ccccccccc}
\hline \hline
ESC & BAC & EMC & ARR(\%) & TR(\%) & SR   & CR    & SoR   & MDD(\%) & Vol(\%) & ENT  & ENB  \\ \hline
 \checkmark    &    &     & 43.88   & 197.88 & 1.90  & 4.89  &  42.20 & 21.70    & 1.12    &  2.56 &  1.23 \\
    &  \checkmark  &     &  45.58   &  208.52 & 1.67 & 4.54  & 36.36 & 24.68   & 1.34    & 2.39 & 1.17 \\
    &    &  \checkmark   & 40.97   & 180.13 &  1.96 & 5.07 & 39.58 & 19.59  & 1.02   & 1.97 & 1.08 \\ \hline
    &  \checkmark   &  \checkmark    & 50.68   & 242.11 & 2.01 & 6.92  & 45.60  & 17.26   & 1.19    & \textcolor{black}{2.61} & \textcolor{black}{1.28} \\
 \checkmark   &    &  \checkmark    & \textcolor{black}{44.57}   & \textcolor{black}{202.17} & 2.24 & 8.78  & 55.01 & 12.02   & 0.94    & 2.72 & 1.39 \\
 \checkmark    &  \checkmark   &     & 59.81   & 308.12 & \textcolor{black}{1.93} &\textcolor{black}{ 5.69 } & \textcolor{black}{40.16} & \textcolor{black}{24.44}   & \textcolor{black}{1.44 }   & 2.91 & 1.42 \\ \hline
 \checkmark    &  \checkmark   &  \checkmark    & \textbf{71.60}    & \textbf{405.34} & \textbf{2.41} & \textbf{11.02} & \textbf{58.00}    & \textbf{8.68}   & \textbf{1.30}     & \textbf{3.13} & \textbf{1.53} \\ \hline \hline
\end{tabular}
}
\end{table*}

\begin{figure*}[htbp]
  \centering
   \includegraphics[width=1\textwidth,keepaspectratio]{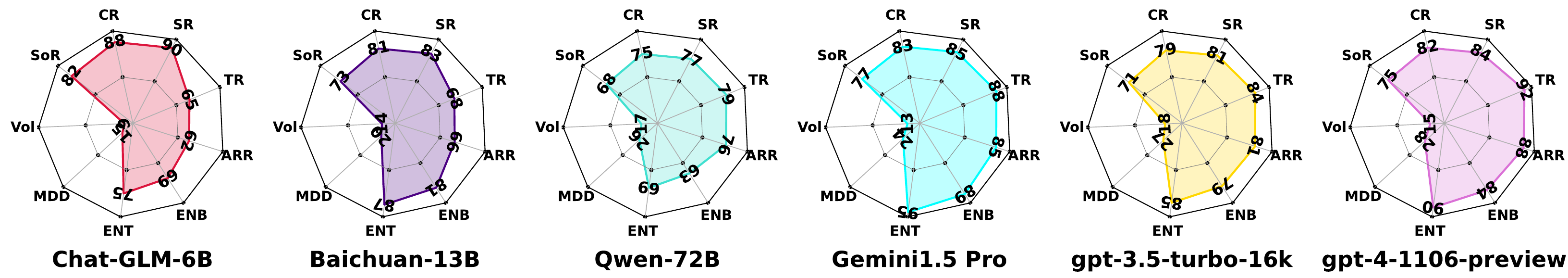}
  \caption{Ablation analysis on several LLM backbones, from open-source to closed-source models.}
  \label{fig:rq3}
\end{figure*}

\subsection{Overall Performance Comparison} 
We selected a diverse range of classical and state-of-the-art baseline models for comparison purposes. These encompass three classical rule-based quantitative investment strategies: MV\cite{YU2011367}, ZMR\cite{eeckhoudt2018dual}, and TSM\cite{moskowitz2012time}; three reinforcement learning-based financial agents: SAC \cite{haarnoja2018soft}, DeepTrader\cite{Wang_Huang_Tu_Zhang_Xu_2021}, and AlphaMix+\cite{sun2023prudexcompass}; and three LLM-based methods: FinGPT\cite{yang2023fingpt}, FinMem\cite{yu2023finmem}, and FinAgent\cite{zhang2024multimodal}. To ensure a fair comparison, we meticulously adhered to the specific requirements of each baseline, providing them with the necessary conditions for optimal performance. 

The experimental results are presented in Table \ref{tab:main}. The following observations can be made: 1) Compared to rule-based strategies, RL-based methods demonstrate a superior capacity to comprehend the intricacies and uncertainties in financial markets . For instance, DeepTrader and AlphaMix+ achieve ARR of 32.78\% and 37.59\%, respectively, which are significantly higher than rule-based models MV (13.03\%), ZMR (-7.25\%), and TSM (19.13\%), indicating the superior learning and decision-making processes of RL-based agents.
2) Leveraging LLMs, as exemplified by FinGPT, FinMem, and FinAgent, enables the formulation of well-informed decisions, surpassing RL-based methods in risk-adjusted metrics such as SR and CR. FinMem achieves an ARR of 47.67\% and a TR of 221.99\%, figures that are substantially higher than RL-based agents, suggesting that LLMs' ability to process and contextualize information leads to more effective investment strategies.

3) Our HedgeAgents exhibits exceptional performance across all metrics, attributable to the following factors:
i) \textbf{Strategic Budget Allocation}.
The dynamic budget allocation plays a pivotal role in optimizing investments across diverse asset categories. This is evidenced by the model achieving the highest ARR of 71.60\% and TR of 405.34\% among all baselines, showcasing the efficacy of strategic budget distribution in enhancing financial outcomes;
ii) \textbf{Sophisticated Risk Management}.
HedgeAgents demonstrate an impressive MDD of 14.21\%, surpassing all other baseline models in risk management.
iii) \textbf{Diversification and Robustness}.
With the highest ENT and ENB, HedgeAgents have demonstrated a diversified investment portfolio that is resilient to the unique risks associated with individual assets. The consistent outperformance of HedgeAgents, as further substantiated by the cumulative returns depicted in Figure \ref{fig:CRAll}. Notably, during May 2022, while most portfolios managed by benchmark models experienced significant losses, HedgeAgents adeptly navigated through this challenging period, demonstrating its superior capabilities.

In summary, these analyses confirm that HedgeAgents has not only surpassed benchmarks in individual metrics but also achieved a harmonious balance of returns, risk management, and diversification. This positions HedgeAgents as a cutting-edge solution, adeptly equipped to operate within complex and dynamic investment environments.

\begin{figure*}[t]
    \centering
    \begin{subfigure}{0.31\textwidth}
        \centering
        \includegraphics[width=\textwidth]{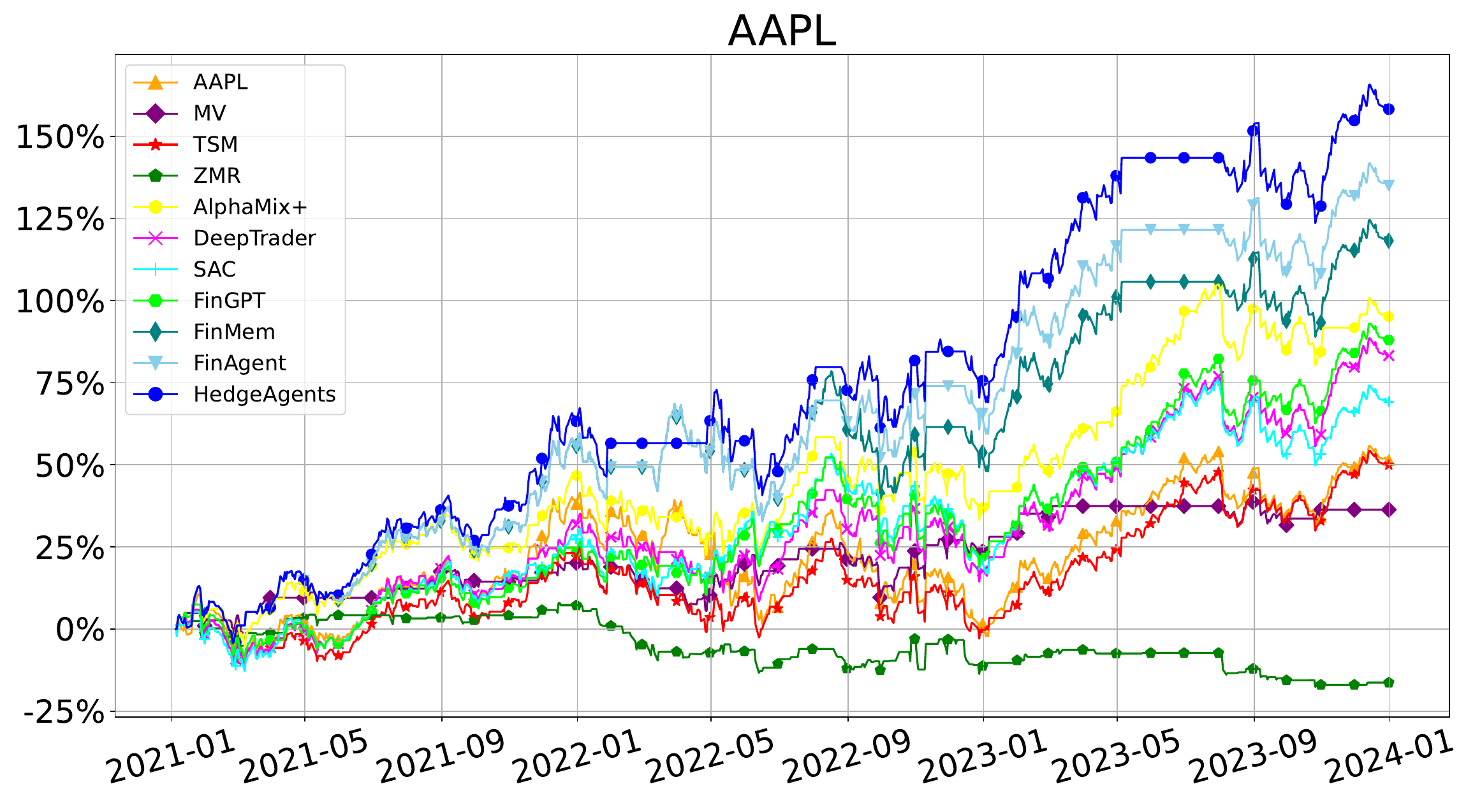}
        \caption{Performance comparison on AAPL stock}
    \end{subfigure}
    \begin{subfigure}{0.31\textwidth}
        \centering
        \includegraphics[width=\textwidth]{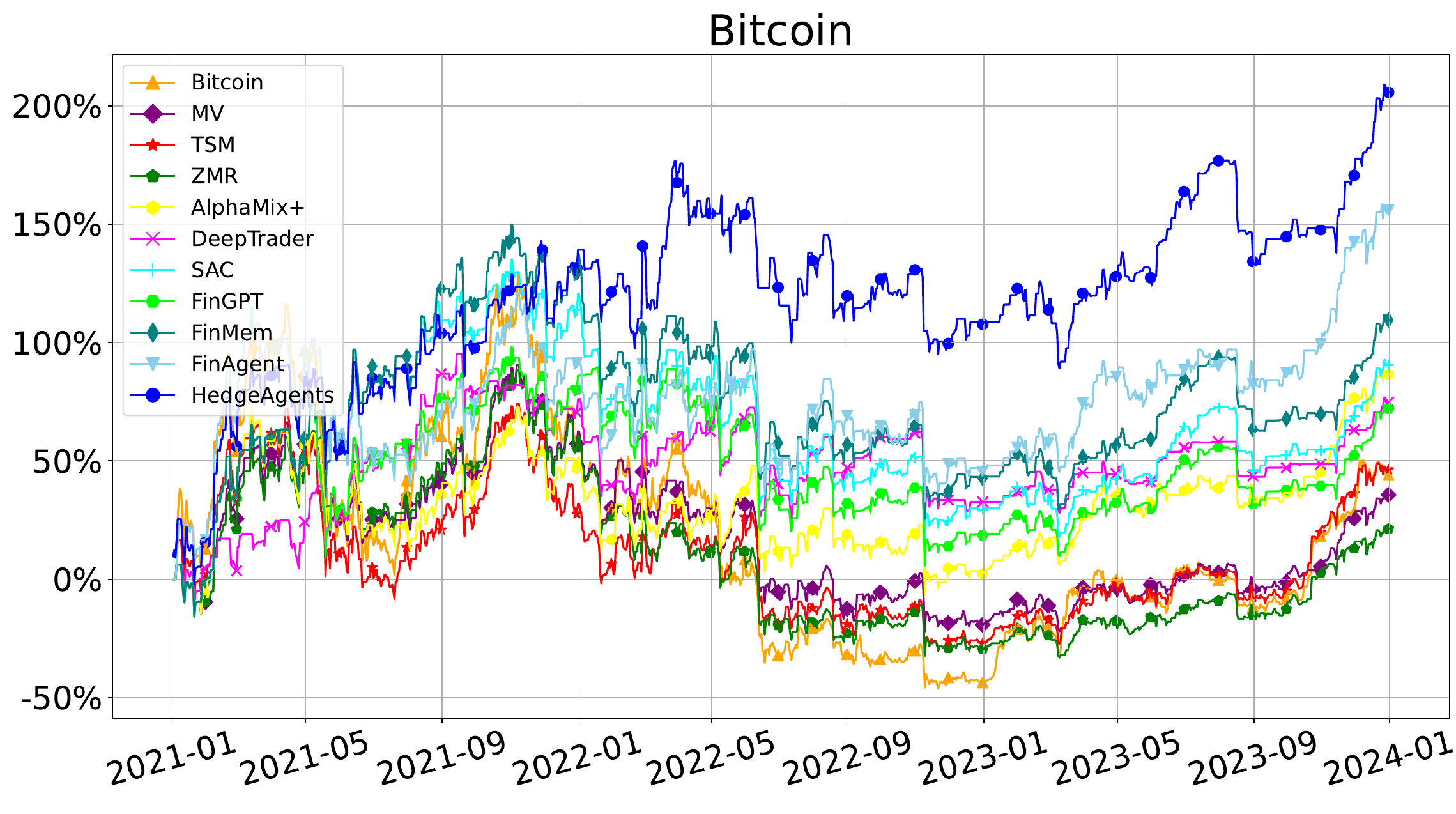}
        \caption{Performance comparison on Bitcoin}
    \end{subfigure}
    \begin{subfigure}{0.31\textwidth}
        \centering
        \includegraphics[width=\textwidth]{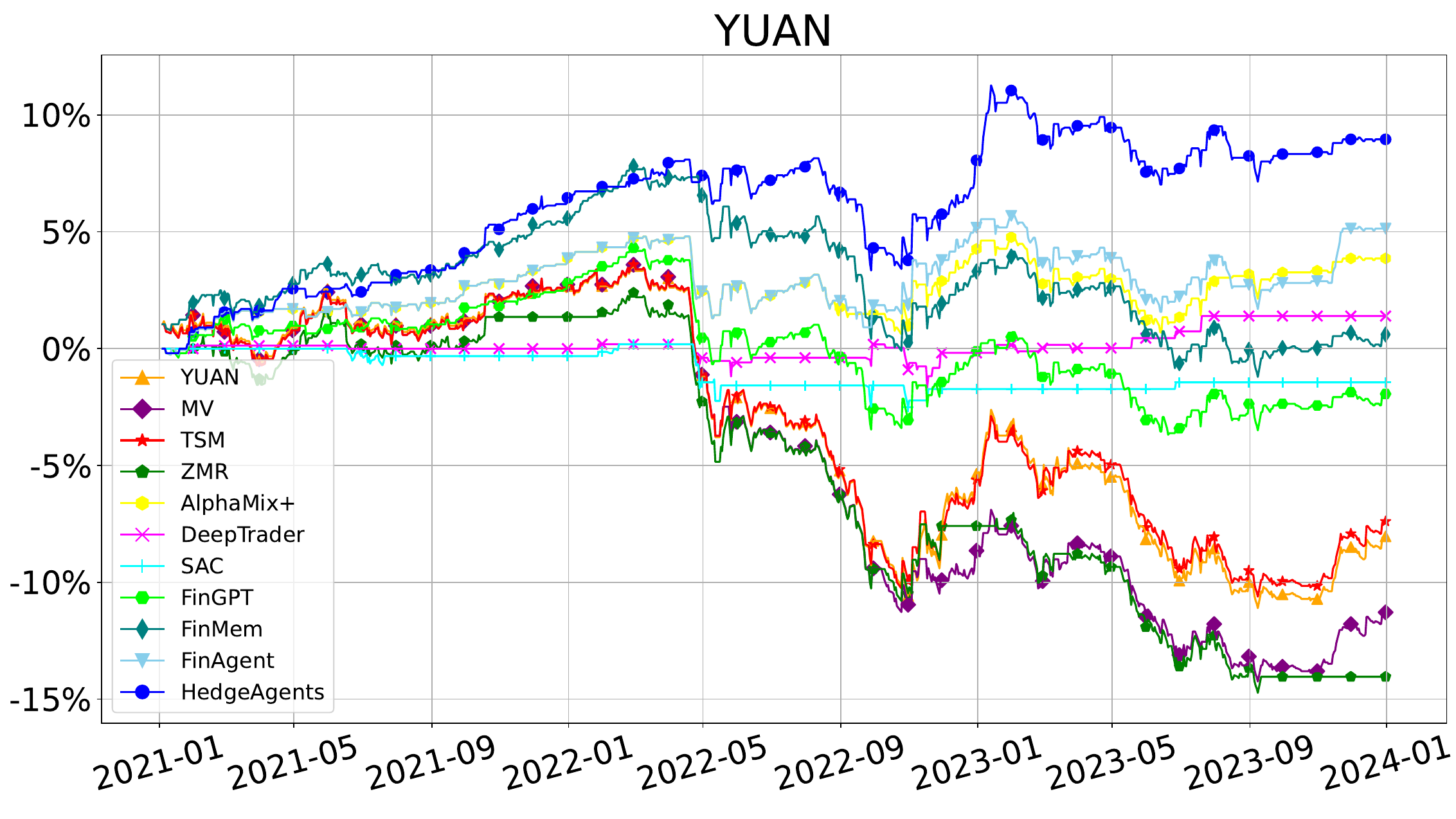}
        \caption{Performance comparison on YUAN}
    \end{subfigure}
    \vspace{-0.3cm}
    \caption{Comparative analysis of cumulative returns for all models in single-asset trading scenarios. These curves show performance differences for AAPL, Bitcoin, and YUAN over a three-year period.}
    \label{fig:single_asset_performance}
    \vspace{-0.5cm}
\end{figure*}

\subsection{Ablation Study}

\subsubsection{Effectiveness of Each Conference}
To comprehensively evaluate the utility of Experience Sharing Conference (ESC), Budget Allocation Conference (BAC), and Extreme Market Conference (EMC) in our proposed multi-agent architecture, we conducted a systematic ablation study. 

The experimental results are shown in Table \ref{tab:rq2}, We have the following observations:
1) The ESC module is crucial for portfolio diversification and hedging tail risks associated with individual assets, which is beneficial for enhancing returns and risk resistance capabilities. When this module is removed, the ENT metric decreases by 16.63\% to 2.61, showing the largest drop; although the annualized return and Sharpe ratio decrease, they remain at a moderate level. 2) The BAC module has the greatest impact on returns. After removing this module, the annualized return drops by 37.75\% to 44.57\%, the lowest among the three single-module experiments; however, when only this module is retained, the annualized return reaches the highest value of 45.58\% for a single module. 3) The EMC module is vital for risk aversion and return robustness. After removing this module, the maximum drawdown rate increases by 71.93\% to 24.44\%, and the Sharpe ratio decreases by 19.86\% to 1.93; however, when only this module is retained, both the maximum drawdown rate and Sharpe ratio achieve the best levels among single modules. 4) The synergistic effect of the three modules significantly impacts the overall performance. Taking the Sharpe ratio as an example, compared to the removal of the other modules alone, the Sharpe ratio of the complete architecture improves by 41.29\%, 60.65\%, and 19.72\%, respectively, highlighting the synergistic effect of each module and ensuring a balance between risk and return. 
This demonstrates the superior performance of our effective multi-agent architecture.

\subsubsection{Effectiveness of LLM Backbone}
To evaluate the performance of different LLMs as the backbone in our HedgeAgents, we selected 6 representative models, including ChatGLM-6B\cite{zeng2022glm}, Baichuan-13B\cite{baichuan_inc_2023}, Qwen-72B\cite{qwen_72b_2023}, Gemini 1.5 Pro\cite{geminiteam2024gemini}, gpt-3.5-turbo-16k\cite{openai_gpt_3_5_turbo}, and gpt-4-1106-preview\cite{openai2024gpt4}. We incorporated each of these models into the HedgeAgents framework and tested their investment performance. 

\textbf{1) Robustness Analysis:} As shown in Figure~\ref{fig:rq3}, despite variations in individual metrics across backbones, the curves shapes are strikingly similar, indicating that our framework is not entirely contingent upon the capabilities of a certain language model.

\textbf{2) Impact of Parameter Size:} Experimental results demonstrate that an increase in parameters contributes to enhancing the investment returns. The gpt-4-1106-preview significantly outperforms others in terms of ARR and TR metrics. Notably, models with more parameters tend to adopt more aggressive investment strategies, which can lead to higher risk. Conversely, ChatGLM-6B achieves the best scores in MDD and SR, showcasing strong risk control capabilities. 

 \textbf{3) Comparison between Open-source and Closed-source:} Closed-source models exhibit excellent performance across all metrics, likely due to their proprietary optimization algorithms and datasets, which provide them with a competitive edge in decision-making capabilities. In the realm of open-source models, Qwen-72B demonstrates performance comparable to the closed-source model.

Therefore, we have selected GPT-4 as the core of our framework. Notably, our system has accumulated a total cost of \$15 over the three years, averaging only 2 cents per day!

\begin{figure*}[t]
  \centering
\includegraphics[width=0.96\textwidth,keepaspectratio]{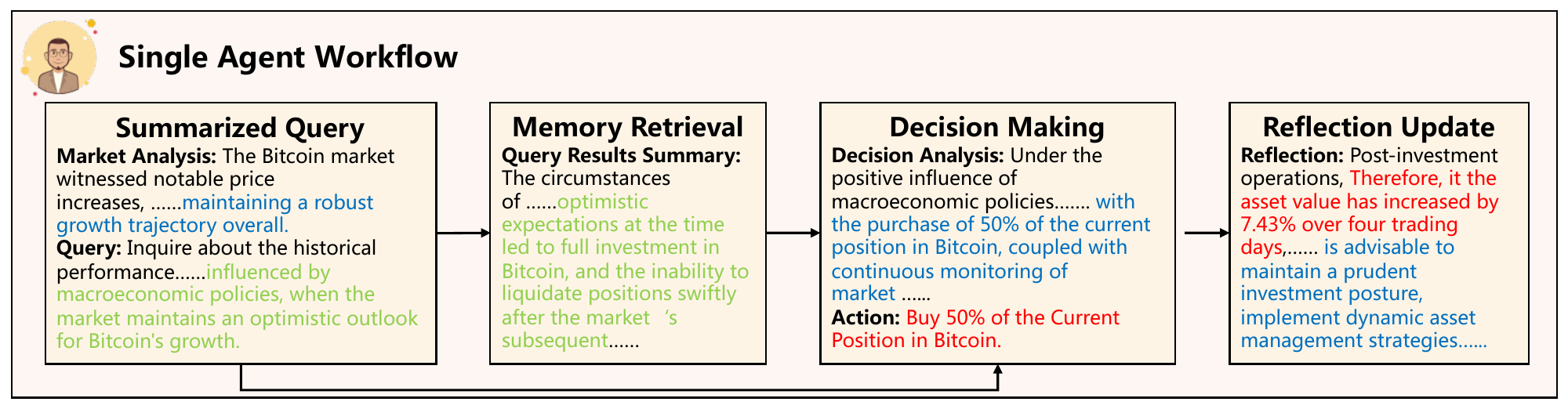}
\vspace{-0.4cm}
  \caption{Workflow of a single agent, taking Bitcoin Analyst Dave as an example. Blue represents the analysis content, green indicates the query process, and red shows the specific decisions and outcomes.}
  \label{fig:sigleagentworkflow}
\end{figure*}

\begin{figure*}[htbp]
  \centering
   \includegraphics[width=0.95\textwidth,keepaspectratio]{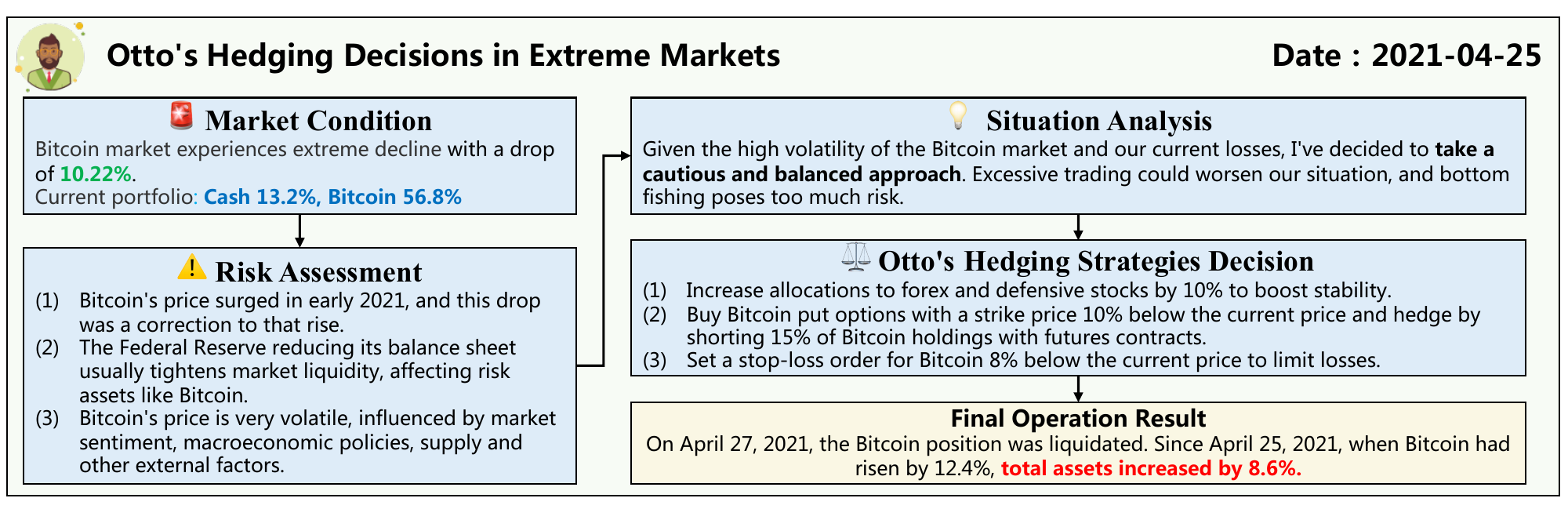}
    \caption{Otto's hedging decisions in extreme markets. Green represents market decline, blue indicates the current portfolio in extreme conditions, and red shows the final execution outcome.}
  \label{fig:ottohedge}
\end{figure*}

\subsection{Single-Asset Performance without Hedging}
In the previous setup, we examined performance in a multi-asset hedging scenario. In this subsection, we will divide our dataset into three categories: Bitcoin-only, AAPL-only (stocks), and YUAN-only (forex). This will allow us to assess the capabilities of all models under single-asset conditions without any hedging strategy.

Figure \ref{fig:single_asset_performance} illustrates the cumulative returns of all models over a three-year period for each single asset. 1) The Bitcoin-only setting showcased HedgeAgents' exceptional management of high-volatility assets, achieving a cumulative return of about 210\%, far exceeding all baseline models. The Bitcoin expert, Dave, excelled in interpreting blockchain data, anticipating regulatory changes, and analyzing global adoption trends, which facilitated stable growth and resilience during market downturns. 2) In the stock market, represented by AAPL, HedgeAgents attained a cumulative return of approximately 160\%, significantly outperforming competitors like FinAgent and FinMem, which recorded returns of about 135\% and 120\%, respectively. The stock expert, Bob, adeptly navigated both bullish and bearish periods, maintaining consistent growth through his skills in analyzing financial statements and predicting market movements. 3) For the forex market, represented by YUAN, HedgeAgents achieved a modest but positive return of around 9\%, surpassing several baselines that showed negative returns. The forex expert, Emily, demonstrated a deep understanding of macroeconomic indicators, geopolitical events, and central bank policies, enabling the model to capitalize on small price movements and achieve profitability.
These results highlight the versatility and robustness of HedgeAgents across various asset classes and market conditions. 


\subsection{Visualization}
To examine our framework's cognitive processes during execution, we have employed visualizations to elucidate the workflow of a single agent and how Otto achieves hedging strategies.

\subsubsection{Visualization of single agent} \label{sec:vsa}
Figure \ref{fig:sigleagentworkflow} illustrates the decision-making process of a single agent, exemplified by Bitcoin Analyst Dave. Dave's workflow begins with a market analysis noting significant price increases and a robust growth trajectory for Bitcoin. This prompts him to query about historical performance under similar optimistic conditions influenced by macroeconomic policies. Through memory retrieval, Dave recalls past instances where optimistic expectations led to full investment in Bitcoin, but also highlights the challenges of swift position liquidation during subsequent market changes. This historical context informs Dave's decision-making phase, where he considers the positive influence of current macroeconomic policies. Balancing optimism with caution, Dave decides to purchase 50\% of the current position in Bitcoin, complemented by continuous market monitoring. This measured approach reflects Dave's synthesis of historical lessons and present market dynamics. The reflection phase reveals a 7.43\% increase in asset value over four trading days, validating the effectiveness of Dave's strategy. Consequently, he advocates for maintaining a prudent investment posture and implementing dynamic asset management strategies. Dave's workflow exemplifies an iterative learning process that combines historical insights with real-time analysis, enhancing decision quality in the volatile market.

\subsubsection{Visualization of hedging expert}
Figure \ref{fig:ottohedge} illustrates the decision-making process of Otto, our hedging expert agent, during a significant Bitcoin market decline of 10.22\%. In this scenario, Otto first assesses the market condition, noting the current portfolio composition of 13.2\% cash and 56.8\% Bitcoin. Otto's risk assessment identifies three key factors: the recent Bitcoin price surge and subsequent correction, the impact of Federal Reserve policies on market liquidity, and Bitcoin's inherent volatility. Based on this analysis, Otto concludes that a cautious and balanced approach is necessary to protect the portfolio. The situation analysis reveals Otto's concern about the high volatility of the Bitcoin market and the current losses, leading to the decision to avoid excessive trading and risky bottom-fishing strategies. In response to these conditions, Otto proposes a comprehensive hedging strategy. This strategy combines portfolio diversification, options trading, and risk management techniques to create a multi-layered hedge against market volatility. By employing this balanced approach, Otto aims to mitigate downside risk while preserving upside potential in adverse market conditions. The effectiveness of Otto's approach is evident in the final operation result, where total assets increased by 8.6\% over two days, despite the initial market turbulence. 

\section{Conclusions}
This paper presents HedgeAgents, an innovative multi-agent system designed to bolster the robustness of financial actions via hedging strategies. Within this crafted framework, we have developed a suite of specialized hedging agents and organized regular conferences to foster collaboration among multiple agents. Through extensive experiments, our framework has consistently demonstrated superior and robust performance. Furthermore, our observations via visualization indicate that our HedgeAgents is capable of generating investment experience in their memories on par with those achieved by human experts. 


\section{Acknowledgments}
This work is supported in part by the National Natural Science Foundation of China (62372187), in part by the National Key Research and Development Program of China (2022YFC3601005) and in part by the Guangdong Provincial Key Laboratory of Human Digital Twin (2022B1212010004).
The authors are highly grateful to the anonymous reviewers for their careful reading and insightful comments.

\clearpage
\newpage
\bibliographystyle{ACM-Reference-Format}
\balance
\bibliography{hedge_www}

\appendix

\begin{table*}[]
\caption{Performance comparison of different LLM as the backbone for HedgeAgents on 9 evaluation metrics. }
\label{tab:llm}
\small
\centering
\normalsize
\setlength{\tabcolsep}{6pt} 
\renewcommand{\arraystretch}{0.99} 
\resizebox{1\textwidth}{!}{ 
\begin{tabular}{cccccccccc}
\hline
\hline
LLM & ARR(\%) & TR(\%) & SR   & CR    & SoR   & MDD(\%) & Vol(\%) & ENT  & ENB           \\ \hline

Chat-GLM-6B        & 49.14         & 231.73          & \textbf{2.88} & \textbf{12.26} & \textbf{66.64} & \textbf{9.19} & \textbf{0.78} & 2.46          & 1.25          \\
Baichuan-13B       & 53.24         & 259.85          & 2.39          & 8.85           & 55.59          & 13.81         & 1.02          & 2.99          & 1.46          \\
Qwen-72B           & 61.34         & 319.99          & 2.16          & 8.01           & 50.21          & 17.44         & 1.29          & 2.35          & 1.18          \\
Gemini1.5 Pro      & 68.61         & 379.37          & 2.49          & 11.48          & 59.94          & 13.12         & 1.21          & \textbf{3.31} & \textbf{1.57} \\
gpt-3.5-turbo-16k  & 65.44         & 352.81          & 2.27          & 8.44           & 53.49          & 17.35         & 1.29          & 2.93          & 1.49          \\
gpt-4-1106-preview & \textbf{71.6} & \textbf{405.34} & 2.41          & 11.02          & 58             & 14.21         & 1.3           & 3.13          & 1.53          \\
\hline
\hline
\end{tabular}
}
\end{table*}

\section{Experiment of Ablation Study} \label{app:ablation}
\subsection{Effectiveness of Each Conference}
\begin{figure}[htbp]
  \centering
   \includegraphics[width=0.49\textwidth,keepaspectratio]{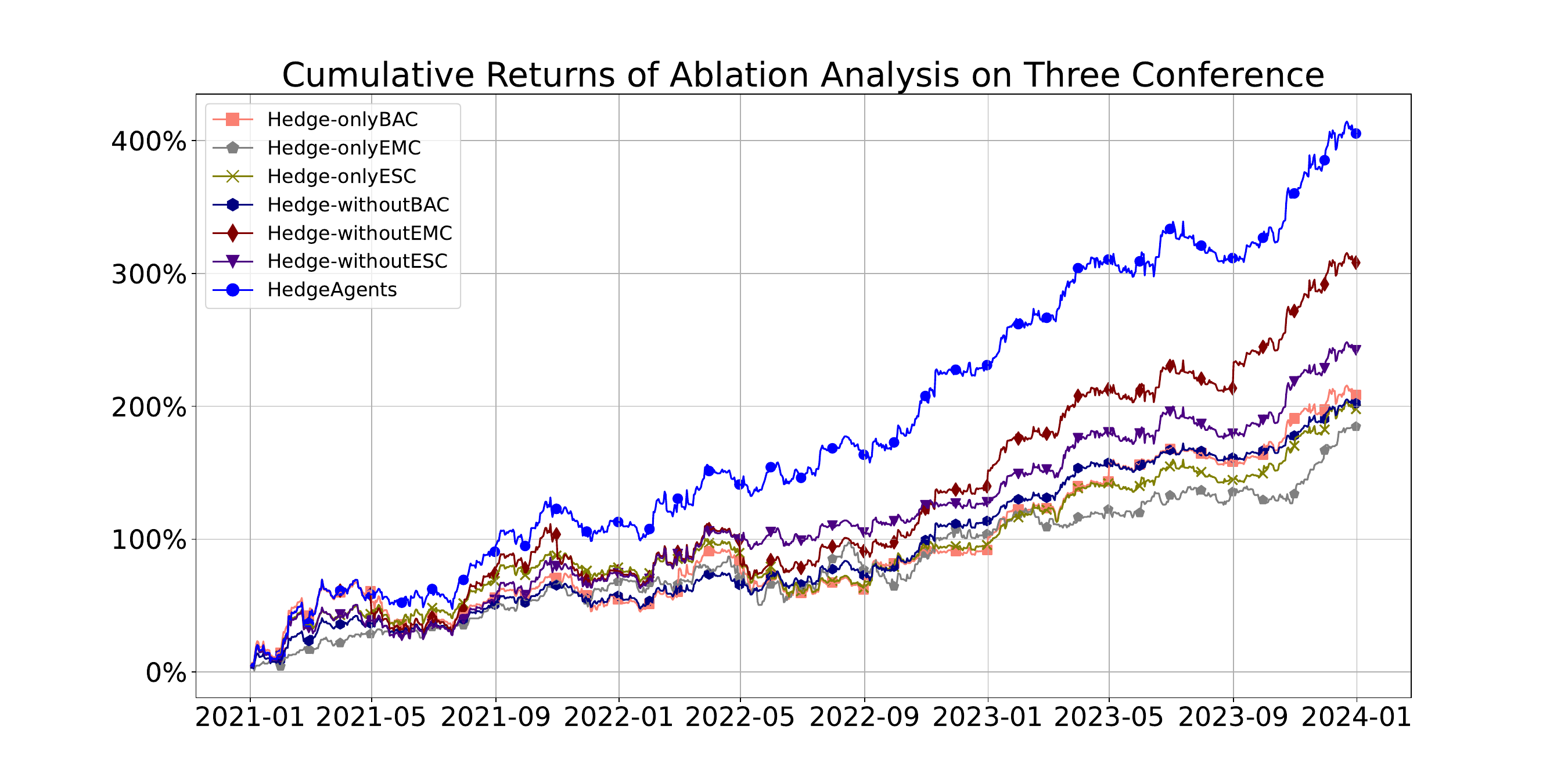}
  \caption{Cumulative Returns of Ablation Analysis on Three Conference}
  \label{fig:CRaba}
\end{figure}

Cumulative returns of ablation analysis on three conference, as shown in Figure \ref{fig:CRaba} .
\subsection{Effectiveness of LLM Backbone}
\begin{figure}[htbp]
  \centering
   \includegraphics[width=0.48\textwidth,keepaspectratio]{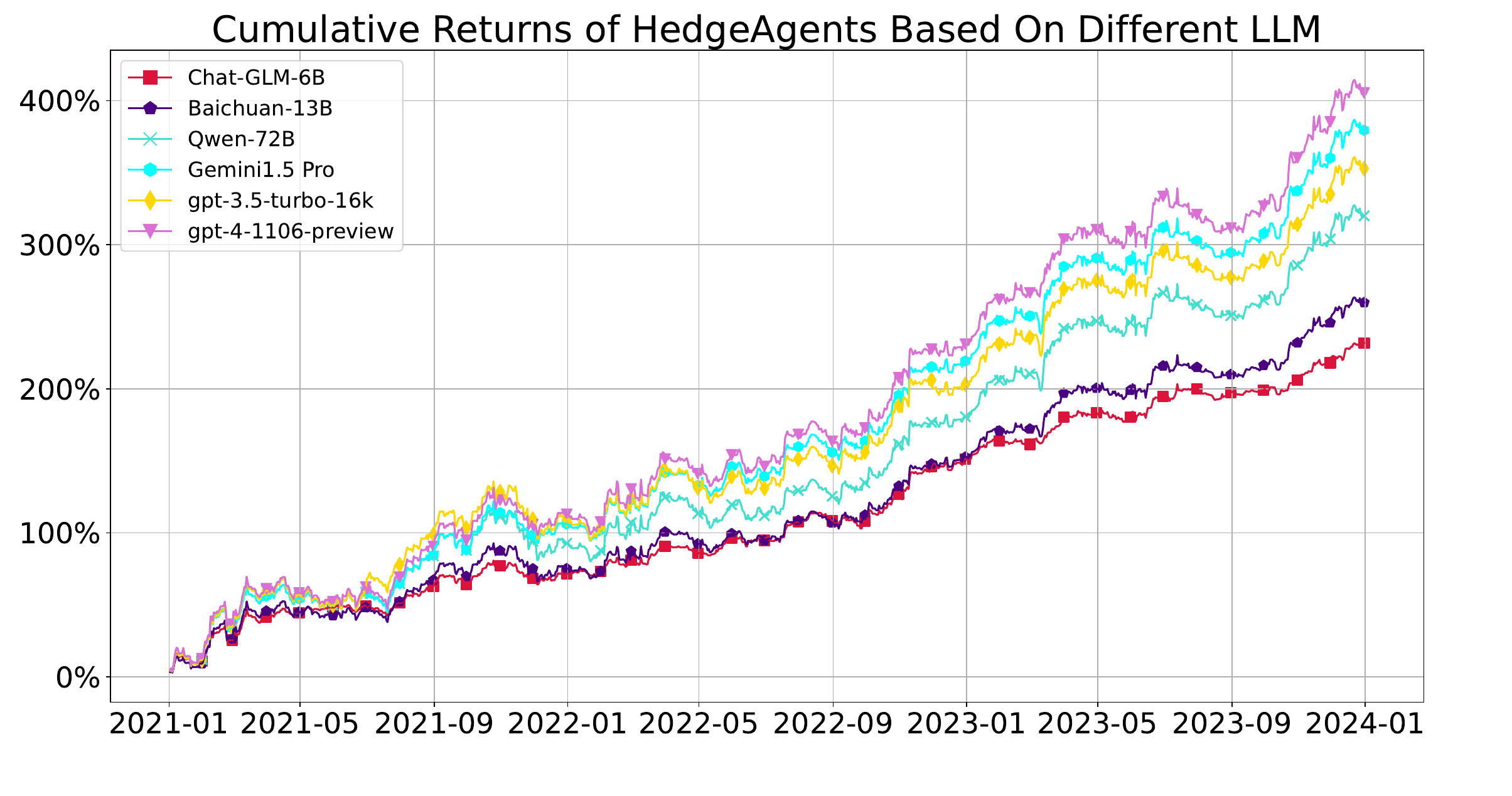}
  \caption{Cumulative Returns of HedgeAgents Based on Different LLM}
  \label{fig:CRabb}
\end{figure}

For using different LLM as the backbone for HedgeAgents, their experimental results are presented in Table \ref{tab:llm}, and the cumulative returns chart is in Figure \ref{fig:CRabb}.

\FloatBarrier

\section{Experiment of Memory Retrieval}
\label{app:memroy}
To quantify the impact of the Memory Retrieval (MR) module on the overall performance of HedgeAgents, we have specifically designed comparative experiments with and without this module.
\begin{figure}[htbp]
  \centering
   \includegraphics[width=0.43\textwidth,keepaspectratio]{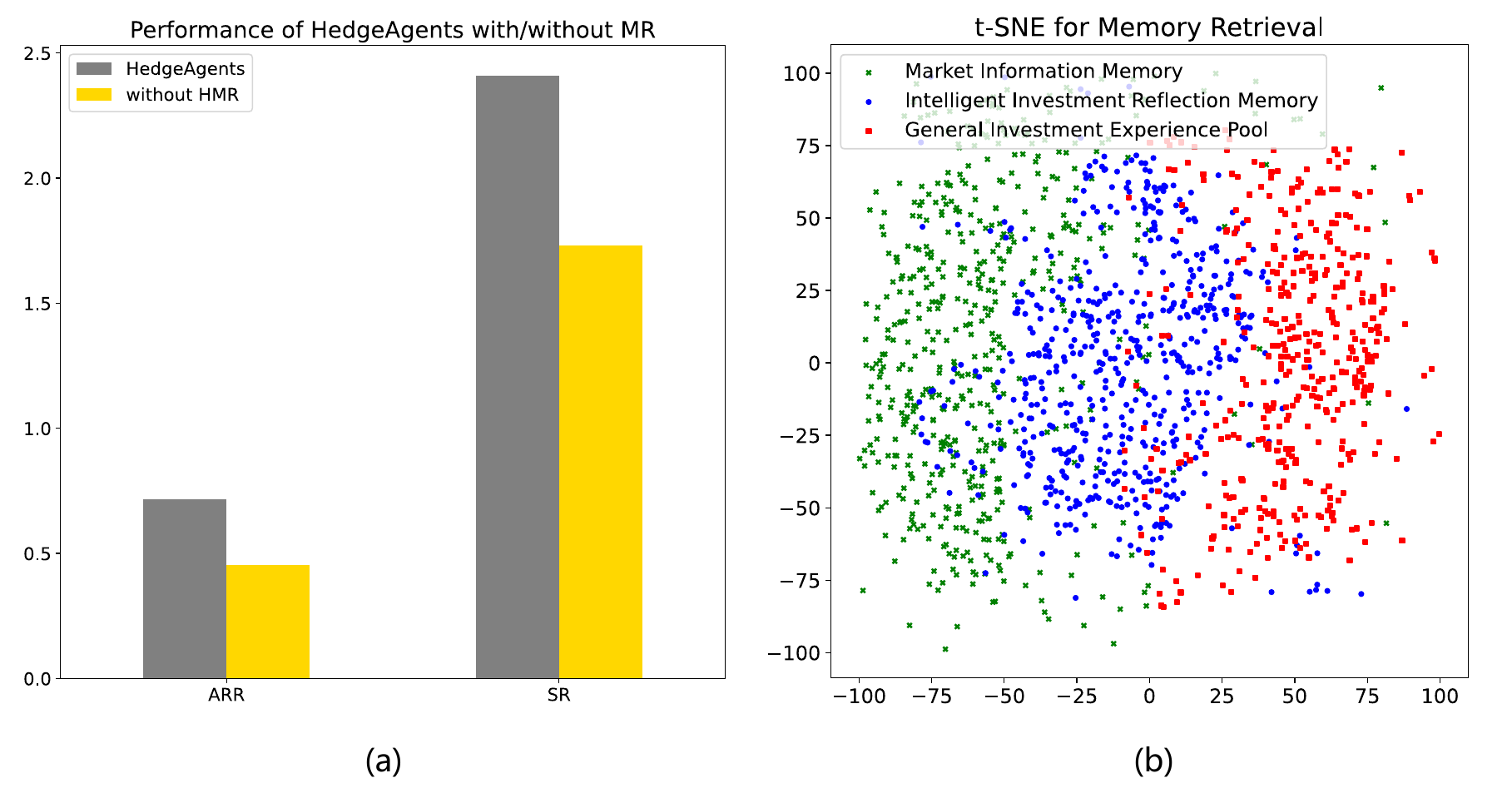}
  \caption{(a) Performance of HedgeAgents with/without MR. (b) Visualization of MR embedding by t-SNE on Dow Jones analyst Bob.}
  \label{fig:Xa}
\end{figure}

As depicted in Figure~\ref{fig:Xa} (a), the Annualized Return Rate (ARR) and Sharpe Ratio (SR) have both significantly increased upon the activation of the MR. Compared to the ARR and SR without the module the enhancements reached 57.71\% and 39.3\%, directly confirming the positive role of the MR module in promoting returns and hedging risks.
To delve deeper into the operational mechanics of the MR module, let's consider the case of DJ30 analyst Bob. We'll illustrate the distribution of various types of memory embeddings when employing this module, as depicted in Figure~\ref{fig:Xa} (b). By employing t-SNE for dimensionality reduction and visualization, we have observed that embeddings derived by LLM from hierarchical memory contents, including Market Information Memory, General Investment Experience Pool, and Intelligent Investment Reflection Memory, showcase a discernible clustering distribution in semantic features. This indicates a certain level of distinctiveness between different categories. Such findings substantiate the efficacy of our Hierarchical Memory Retrieval approach in effectively capturing the semantic attributes of memory content, thereby furnishing the agent with valuable contextual memory support.
\section{Experiment of Temporal Isolation}
\begin{figure}[H]
    \centering
    \includegraphics[width=0.45\textwidth]{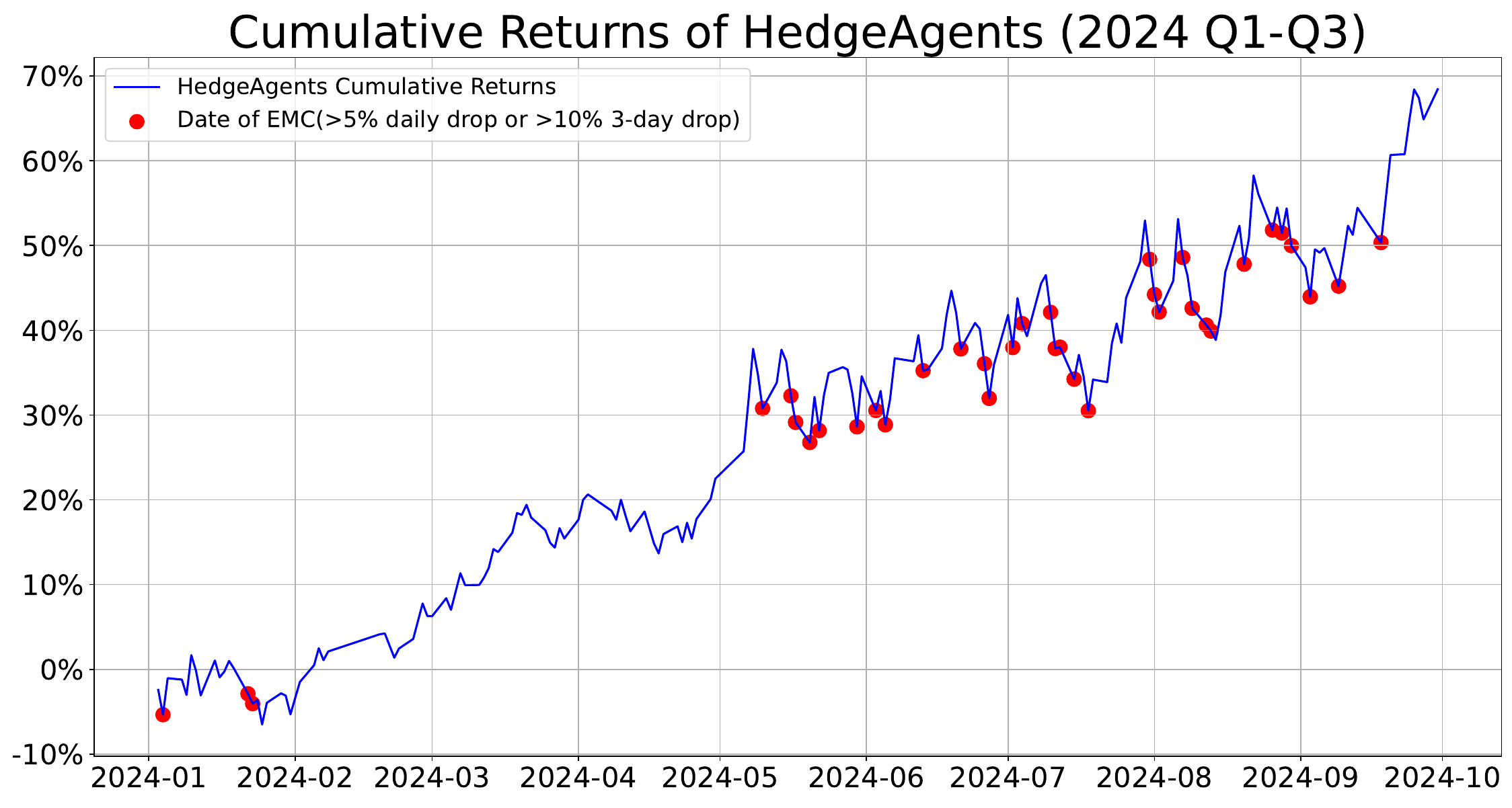}
    \caption{Cumulative Returns of HedgeAgents. Red markers indicate dates of EMC meetings, which occurred 36 times.}
    \label{fig:hedgeagents_performance}
\end{figure}

To minimize the risk of inadvertent information leakage, we designed context-rich prompts incorporating real-time conditions, market variables, and asset-specific data.
Figure \ref{fig:hedgeagents_performance} shows the performance of HedgeAgents during Q1-Q3 of 2024, demonstrating its ability to adapt to rapidly changing market with high predictive accuracy and robustness. The model achieved a Total Return of 68.44\%, a Sharpe Ratio of 2.1. These results validate the model’s robustness and confirm the effectiveness of our information leakage mitigation strategies.

\end{document}